# Integrated Self-Organized Traffic Light Controllers for Signalized Intersections

Maythem K. Abbas[1], Mohd Noh Karsiti[2], Madzlan Napiah[3], Samir Brahim[4]

**Abstract** – *Detecting emergency vehicles arrival on roads has been the focus for many researchers. It is quite important to detect the emergency vehicles (e.g; ambulance) arrival to traffic light to give the green light for it to pass through. Many researchers have suggested and patented emergency vehicles detection systems however, according to our knowledge, none of them considered solving the effect of giving extra green time to a road while the queues are being built on others. This paper considers the problem of finding a better traffic light phase plan to stabilize/recover the situation at an effected intersection after solving an emergency vehicle existence. A hardware setup and a novel messaging protocol have been suggested to be set on roads and vehicles to collect roads real time data. In addition, a novel decision making protocol has been created to make the use of the collected data for making a better traffic light phase plan for an intersection. The phase plan has two main decisions to be made; which light has a higher priority to be green in the next phase, and how long the green phase should be. After simulating the proposed system using our customized simulator written in Matlab programing language and comparing its performance with other related works, significant enhancements have been observed in terms of stabilizing the queue lengths at an intersection after solving an emergency case.*

## I. Introduction

When an on-duty emergency vehicle arrives to a signalized intersection and need to cross the intersection quickly, traffic light controller should be able to detect its arrival earlier so it can arrange a suitable phase plan to pass the emergency vehicle through. In addition, the traffic light controller needs to monitor the intersection's legs after passing the emergency vehicle through so it can recover back the normal situation for the intersection.

Since the early stages of traffic lights deployment on streets in the 1920s, many researchers have tried to solve the problem of detecting emergency vehicles arrivals to a traffic light signal to let them pass through. Looking back to 1990, a traffic light preemption system have been proposed by two researchers from USA, [8]. They have suggested setting a directional antenna on emergency vehicle to send its identification, priority, and direction to a preemption system set at the signalized intersection. The main aim of their work is to detect and identify the emergency vehicles to give them green lights to pass through. They have ignored how to recover the effect of giving red lights to the rest of the intersection's traffic lights when passing through the Emergency vehicle.

In [9], another try to solve the emergency vehicles existence by setting a device on the vehicle sending a radio signal to the signalized intersection it is approaching to control it remotely and changes the traffic light into green. Again, the same downside appeared when [9] solved the emergency vehicles detection while ignoring the intersection's situation recovery as it appeared with all the works [7], [10], [5], [3], [2], and [4].

The main objective of this study is to propose an emergency vehicle detection system using VANET (Vehicular Ad-Hoc Networks) and an algorithm for self-organizing traffic light controller as a part of Dynamic Traffic-Light's Phase Plan Protocol (DT3P).

This study considers the problem of finding a better traffic light phase plan to stabilize/recover the situation at an effected intersection after solving an emergency vehicle existence.

The main focus in this study is to update the existing technology for emergency vehicles detection and propose a novel algorithm for recovering the effect of giving green time to one of

M. K. Abbas, M. N. Karsiti, M. Napiah, S. Brahim

the intersection's legs while the queues are being built up at the legs with the red lights.

Detecting the existence of emergency vehicles had become quite simple when using VANET. This paper focuses more on the intersection's situation recovery system. Two works have been chosen to compare the results with; [1] and [6], as they both focus on optimizing intersections situations. Both of the works have suggested using the same hardware as shown in **Fig.1**. They have suggested setting up a sensor belt on each road to count the arriving vehicles for queuing but the difference is the decision making algorithms to answer the two questions; Which leg has higher priority to become green next phase and how long the next phase should be.

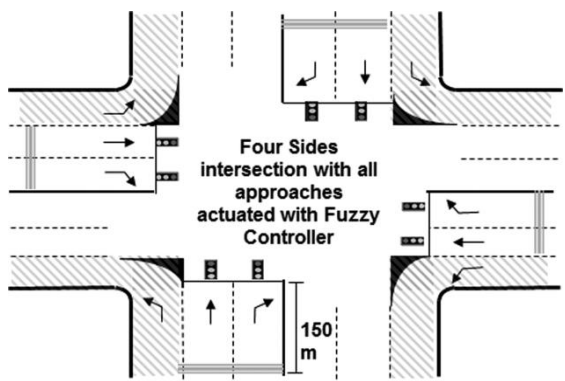

Fig. 1. hardware setup for both of [1] and [6].

Like in [1] and [6], our approach using a standard four legs intersection as shown in **Fig.2**. Each leg has three directions and each of them represented by a single lane. The left direction of all the legs are considered slip lane (signalized lane), this is why it would be neglected when simulating the results.

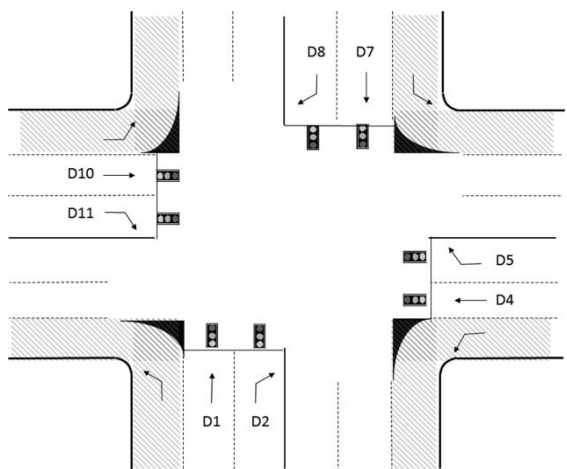

Fig. 2. Standard Four Legs Intersection

As in [1], the priority of each approach was determined by summing the total number of vehicles queuing and that can peak at 50 vehicles per approach. In [6], they looked to the phase priority almost the same as in [1] but the maximum queue length is 60 vehicle per approach.

Both of the above works have limitations in terms of choosing the phase with the highest priority. Both of [1] and [6] works, have a fixed maximum queue length, 50 vehicles per approach and 60 vehicles per phase respectively. At the same time, they both neglected the first vehicle waiting time, and that would let the approaches with the very low volumes to wait for very long time.

Another point of highlight In [1] work, that the total number of phase options are four only which would reduce the chances to choose the most optimum phase sequence. While our approach increases the chance to enhance the decision making quality by widen the range of phase choices up to 12 choices.

By overcoming the downsides of the other works and adding some new features, our approach is expected to offer a better signalised intersection control mechanism.

## II. Road's Data Collection

The hardware setup suggested in **Fig.3**, would collect the road status from the road and the nearby roads. Those collected data would be sent to Traffic Light Controller (TLC) which is placed at the intersection to help making an optimum decision about the next green phase.

The setup of the Road's Data Collection System (RDCS) can be described as a set of road side equipments (RSE) are sited beside the road, each connected to a sensor belt which is set on or under the road pavement. The job of the sensors belt depends on its position, as the job of the belt at the road's main entrance is to detect the total number of vehicles arriving each lane, that we named it as Load Adder. While the traffic light stopping line's belt job is to detect the number of vehicles leaving each lane, as it's named Load Subtract. Six meters before the stopping line, another belt is positioned to confirm the first vehicle arrival on each lane. The last type of belts is the Load Estimators. The main duty of the load estimators is to detect the queue length on each lane. The number of the Load Estimators depends on the length of the street, as they are separated by 150 meters. An important point to highlight about the Load estimators that only one load estimator is doing the queue length

M. K. Abbas, M. N. Karsiti, M. Napiah, S. Brahim

detection per lane at a time, and that depends of the queue length. For example; considering the case when the queue length on the uppers lane, **in Fig.3**, is less than 30 vehicles while the queue length of the middle lane has more than 30 vehicles, then first Load estimator would detect the queue length for the upper lane, while the second load estimator would do for the middle lane. The coordination of the belts work is the duty of the RSEs.

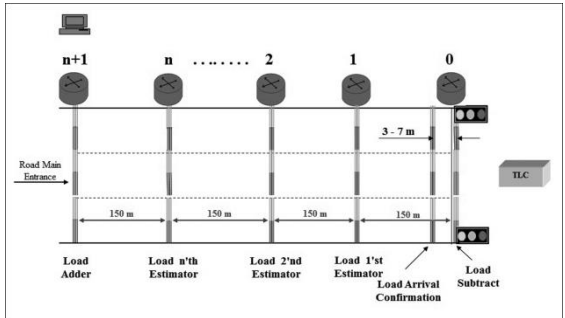

Fig. 3. Road data collection system's hardware setup

The arrival of any vehicle to a road's entrance would start a side to side communication between the vehicles and the road side equipments which would help in the emergency vehicles arrival detection. As soon as the hand-shaking completes between the two sides, a report would sent by the vehicle to the RSE holding some information about the vehicle, including the type of the vehicle and its ON-DUTY flag (0 = Idle, 1 = On-Duty).

## III. Traffic Light Phase Decision Making

After collecting the data from along the roads of an intersection, they would be compiled into 8 variables per direction. Finally, those direction's eight variables would be transferred to its last destination; The Traffic Light Controller of that intersection where the decision is being made. See **Fig.4**.

The traffic light Controller would have 8 sets of variables; each set represents one flow direction. Each set contains 8 values those represent the road's latest status. The 8 by 8 data matrix would be used to make two decisions. The first decision would be which pair of flow directions would be green next phase, While the second decision to be made by the Traffic Light controller would be next green phase time.

### III.1. Green Lights Priority

DT3P considers the priority of each lane then the priority of each phase. After collecting the variables representing the road condition at a point of time T, the load of each direction ( $i$ , Where I is the Direction index number = $\{1,2,3,4,5,6,7,8\}$ and $i \in I$) would be calculated within the TLC using those variables in addition to the static values configured when installing the system over the road. See Fig.5.

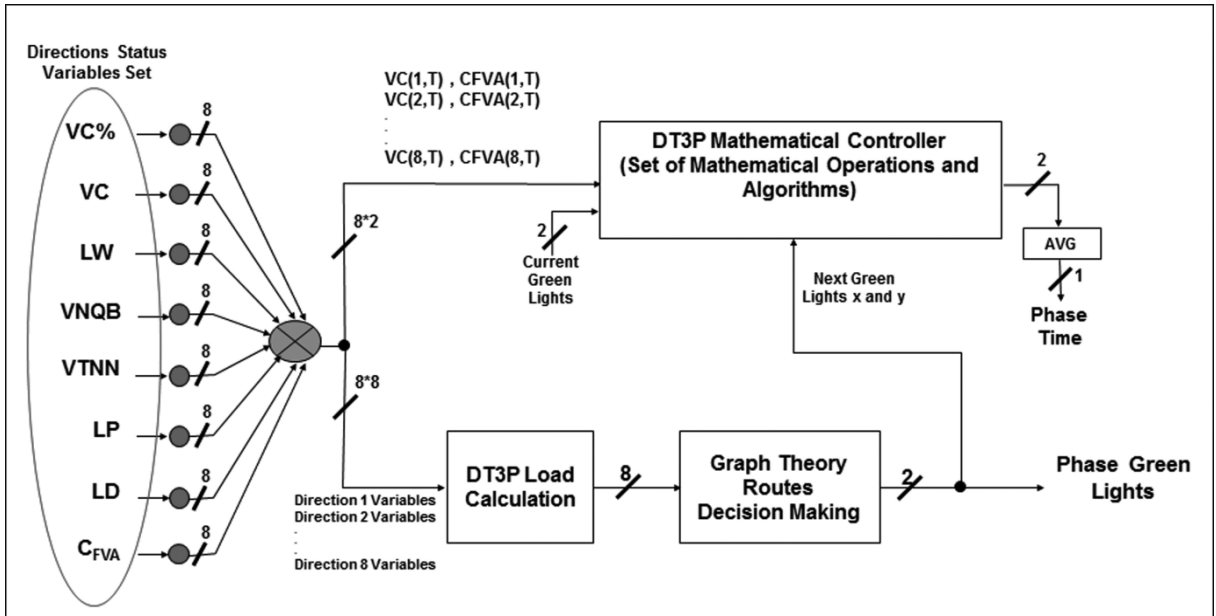

Fig. 4. Traffic light Controller with DT3P Mathematical Algorithm Core

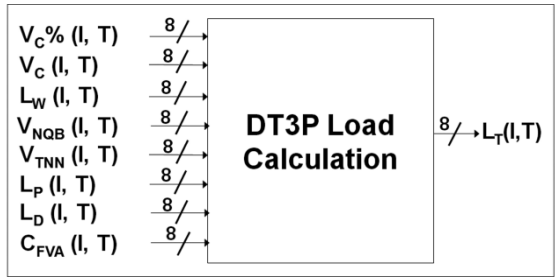

Fig. 5. DT3P Load calculation stage

The factors of variables (Non-Unit values) used to calculate each direction's load would be defined in this section. Starting with $V_C(i,T)$ which can be defined as the Vehicles Count confirmed to be within the Queuing area of direction i at the point of time T. the First vehicle arrival confirmation Flag on direction i at the point of time T Abbreviated as $C_{FVA}(i,T)$. Whereas $V_C\%(i,T)$ stands for how much percent of the vehicles first queuing area of the direction i's road is occupied at the point of time T. When a first vehicle arrives to a red traffic light, a timing counter starts counting up the Waiting time ($L_W(i,T)$) for the first vehicle in the queue of direction i's road at the point of time T. For detecting the emergency vehicles existence, two variables are being collected; vehicle's priority $L_P(i,T)$ and the flag $L_D(i,T)$ for the special vehicle (driving on direction i at the point of time T) whether it's on Duty ($L_D=1$) or not ($L_D=0$). Two variables are being collected for integration purpose, those are $V_{NQB}(i,T)$ (Vehicles Total Number Queuing on the Back-road traffic lights those leading to the direction i's road at the point of time T) and $V_{TNN}(i,T)$ (Vehicles Total Number instantly available (at the point of time T) on the Next road receiving vehicles coming from the direction i's road).

Eventually, using the above collected variables, the load of each direction would be calculated as illustrated in **Fig. A1,** then sent to the Graph Theory Routes Decision Making (Within the TLC) for the next phase green lights decision making, see **Fig.7**.

In [1] work, the total number of phase options are four only which would reduce the chances to choose the most optimum phase sequence. DT3P solved this problem as it would look to each direction at the intersection as a node in a graph, See **Fig.8**. Each node has four relations with the four nodes intersecting with its direction, those called as adjacent nodes.

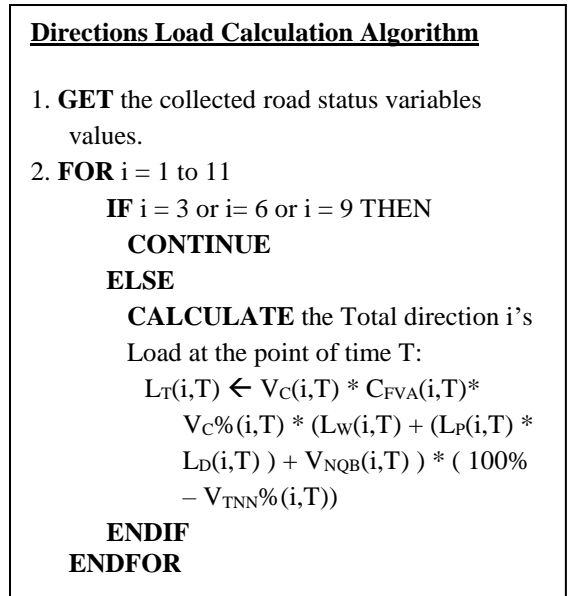

Fig. 6. Directions load calculation algorithm

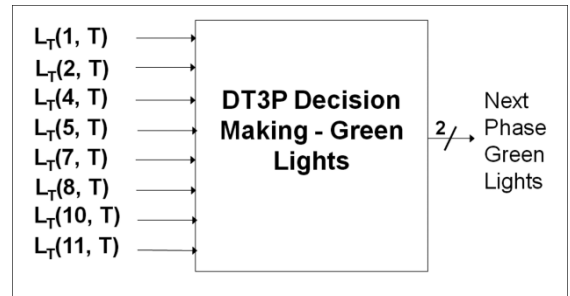

Fig. 7. DT3P Next Phase Green Lights Decision Maker black box

When the time comes to choose which two directions should become green next. Two initialization steps would be implemented. The first step is to set the list of adjacent nodes for each node on the graph. See Fig. A2. Second initialization step would be assigning the values of the calculated loads of each direction to a node on the graph in **Fig.8.** The process of determining the next green lights would start firstly with a one-to-all pairing operation happens between the elements of the first current green's adjacent nodes and the elements of the second current green adjacent nodes to get a list of 16 combinations. Secondly, an elimination operation would happen to remove any paired-to-self combinations (e.g.; Node H is paired to itself). Thirdly, another elimination operation lunched to remove the intersected paired nodes (e.g.; if AC was a paired combination, and A intersects with C, then AC would be removed from the list), leaving the list of the available phase combinations.

M. K. Abbas, M. N. Karsiti, M. Napiah, S. Brahim

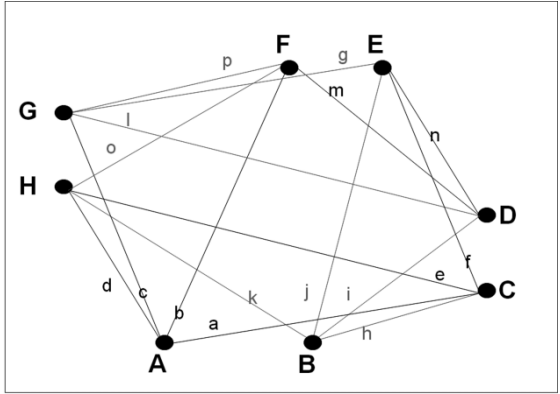

Fig. 8. DT3P Phase Transition Map

**Next Green Phase Lights Algorithm**
1. **INIT** Adjacent Nodes List of:
   Node A ← C,F,G,H
   Node B ← C,D,E,H
   Node C ← A,B,E,H
   Node D ← B,E,F,G
   Node E ← B,C,D,G
   Node F ← A,D,G,H
   Node G ← A,D,E,F
   Node H ← A,B,C,F

2. **SET** A ← $L_T(1,T)$, B ← $L_T(2,T)$, C ← $L_T(4,T)$, D ← $L_T(5,T)$, E ← $L_T(7,T)$, F ← $L_T(8,T)$, G ← $L_T(10,T)$, H ← $L_T(11,T)$

3. **DETERMINE** the available unit-to-unit pairs from the two currently green's adjacent nodes lists.

4. **ELEMINATE** the pair-to-itself combinations.

5. **ELEMINATE** the intersected (unavailable) pairs.

6. **DECENDING SORT** the rest of the pairs in the list.

7. **SET** the first pair two elements on the list as the next phase green lights.

Fig. 9. DT3P Next Phase Green Lights Decision Making algorithm

Finally, the rest of the combinations would be sorted according to their loads summations in descending order. The first phase Combination from the list would be the next green phase.
example: Current green phase is AB

**First Step:** Adjacent List of Node A: C,F,G,H
Adjacent List of Node B: C,D,E,H
Pairs List:

CC,CD,CE,CH,FC,FD,FE,FH,GC,GD,GE,GH,HC,HD,HE,HH

**Second Step:** Paired-to-Itself elimination
Pairs List:

**CC**,CD,CE,CH,FC,FD,FE,FH,GC,GD,GE,GH,HC,HD,HE,**HH**
New Pairs List:
CD,CE,CH,FC,FD,FE,FH,GC,GD,GE,GH,HC,HD,HE

**Third Step:** Intersected pairs elimination
Pairs List:

CD,**CE**,**CH**,FC,**FD**,FE,**FH**,GC,**GD**,**GE**,GH,**HC**,HD,HE
New Pairs List:
CD,FC,FE,GC,GH,HD,HE

**Last Step:** Sorting the pairs list in descend order (Assuming that the loads summation of Nodes C and G are the biggest) then:
Pairs list after sorting: GC, CD,FE,FC,HD,GH,HE
Then the next green phase is GC.

To change a phase, both of the two currently green nodes check their Adjacency List to find the two maximum weighted nodes which are not adjacent to act as the next green phase.

*III.2. Green Phase Time*

Choosing the next green lights is one of the two phase decisions to be made. Now the second decision to be determined is how long the green phase should be.

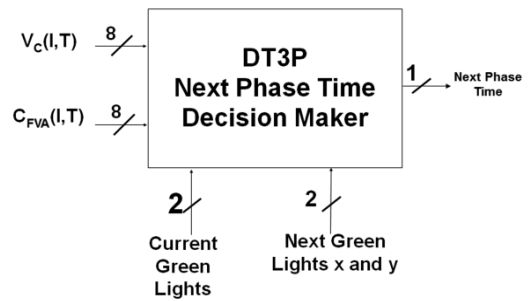

Fig. 10. DT3P's Next Phase Time Decision Maker Block Diagram

The condition of each direction at intersections would be represented by 8 values. Only two values ($V_C$ and $C_{FVA}$) among those 8 values would be sent as inputs to the Phase time decision maker, See

**Fig.10**, In addition to the indexes of two next green lights and the two currently green lights, as we can see in **Fig.10.**

Fig.11 represents the DT3P Mathematical Method for determining the next phase time as blocks diagram. At the point of time T, Twenty values would arrive the DT3P Mathematical controller. Eight of them represent the number of vehicles ($V_C(I,T)$) been confirmed to be arrived at the queuing area of the lane indexed I, where I = {1,2,3,4,5,6,7,8}. While the second eight inputs represent the first vehicle arrival confirmation flag ($C_{FVA}(i,T)$) for each direction i, i ϵ I. The last four inputs to the Next Phase Time decision maker are divided into two sets, the first set carries the two index numbers of the current green directions, and the second set of values holds the two index numbers of the next phase elected directions to be green.

The first step the controller does is to confirm that there is at least one vehicle (at each of the 8 directions) queuing and waiting for the traffic light to become green. If there is no vehicle already arrived at the traffic light of the direction i, then the $V_C(i,T)$ value would be neglected, otherwise it would be passed to the next level as $V_C'(i,T)$. This operation is to be implemented inside a Multiplier that receives 2 sets of 8 values each; $V_C(I,T)$ and $C_{FVA}(I,T)$. Each element in $V_C(I,T)$ set would be multiplied by its equivalent element in $C_{FVA}(I,T)$.

The second stage is to find out which two subsets of the inputs $V_C'(i,T)$ are adjacent to the two chosen green lights and if any of the subsets members are currently green or not. The reason behind multiplying the $V_C'(i,T)$ by the Adjacency flag is to make sure that we are choosing the correct mates for more accurate time decision, as illustrated later in this section. While multiplying by the " Is i NOT currently green?" flag ($G_{Cx}$) is to make sure to not include the currently green lights into the consideration when deciding the next phase time as they are not among the next phase green lights.

The main steps to be done by the mathematical algorithm are shown in **Fig.12**.
  - Get the queues length for each of the eight lanes ($V_C(1,T)$ ..., $V_C(8,T)$).
  - Get which two lanes will be green the coming phase ($G_{N1}$ and $G_{N2}$).
  - Sum the queue lengths of each of the two elected lanes with the queue lengths of its crossing lanes those which are not flagged among the current green lights.
  - Get the summation of the Current Green Light flags for each of the crossing lanes for the elected lanes including themselves.
  - Find the ratio of each of the elected directions to become green next phase to the total queue length with its crossings. As in step 5 in Fig.12.
  - Multiply the results gotten from the previous step by summation of the current green flags summation and by the time for the single green phase (for example; the basic green time for a traffic light is 30 Seconds) by the number of legs of that intersection. See step 6 in **Fig.12**.

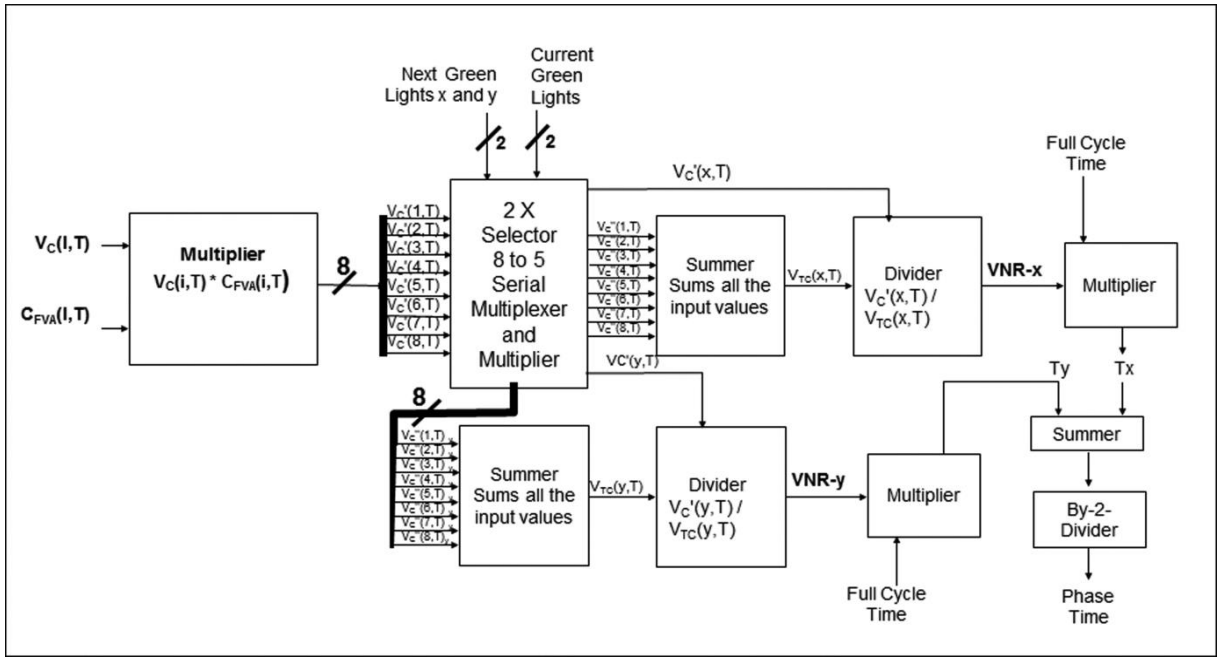

Fig. 11. The DT3P Mathematical Algorithm Core Internal Architecture

- Finally, Find the [Maximum, Average, or Minimum] Time among the two results. If the main purpose was to reduce the overall wasted time at the intersection, then choosing the Minimum Phase time would be the best choice. While if the main purpose was to reduce the queue lengths to minimum, then choosing the maximum required phase time would be the best option. But if the main aim was to maintain both the wasted time and the queue lengths, then it would be better to average both results to get the next phase green time as has been done in step 7 in **Fig.12**.

The second decision has been made and till this moment DT3P hypothesis is just a theory. Next we would like to examine how effective the proposed DT3P to recover the situation at a signalised intersection which have just solved an emergency vehicle existence.

## IV. Simulation Setup

To check how efficient the proposed methodology in this paper compared to other works, a set of case studies have been implemented using our customized simulator which has been validated using ASidra Intersections simulator. The customized simulator offers the ability to write a customized traffic light controller algorithm, unlike ASidra Intersections simulator which does not permit to change the controller algorithm.

The Experiment was designed to work on a standard four legs intersection, as shown in Fig.2. The first two queues (D1 and D2) are empty as they have been given enough green time to release all the queued vehicles including the emergency case. The fourth and the fifth directions have 100 vehicles queuing one each, the seventh and the eighth directions have 75 vehicles queuing, while the tenth and the eleventh directions initially have 50 vehicles each.

---

**Next Green Phase Time Algorithm**

1. **GET** each direction's queue length ($V_C$), the first vehicle's arrival flag ($C_{FVA}$), The Two Currently Green directions IDs ($G_{C1}$, $G_{C2}$), The selected next phase two green lights IDs ($G_{N1}, G_{N2}$), and The standard Full Cycle Time.

2. **DETERMINE** which directions queues are confirmed to be arrived:
   $V_C'(i,T) \leftarrow V_C(i,T) * C_{FVA}(i,T)$

3. **DETERMINE**, for each direction, which of its adjacent queues should be considered in calculating the next phase time while setting the rest (The Non-Adjacent or currently green) of them into zero.
   $V_{CGN1}''(i,T) \leftarrow V_C'(i,T) * $ *(Is Node i adjacent to $G_{N1}$) * (i~=$G_{C1}$ and i~= $G_{C2}$)*
   $V_{CGN2}''(i,T) \leftarrow V_C'(i,T) * $ *(Is Node i adjacent to $G_{N2}$) * (i~=$G_{C1}$ and i~= $G_{C2}$)*

4. **CALCULATE**, for each of the two directions, the summation of the results in step 3.
   $V_{CT}(G_{N1},T) \leftarrow $ *Sum ($V_{CGN1}''(i,T)$) for i = 1 to 11*
   $V_{CT}(G_{N2},T) \leftarrow $ *Sum ($V_{CGN2}''(i,T)$) for i = 1 to 11*

5. **CALCULATE**, for each of the two directions, the division of the chosen direction's queue length over the total summation found for that direction in step 4.
   $VNR_{GN1} \leftarrow V_C(G_{N1},T) / V_{CT}(G_{N1},T)$
   $VNR_{GN2} \leftarrow V_C(G_{N2},T) / V_{CT}(G_{N2},T)$

6. **CALCULATE**, for each of the two directions, How much present of the full cycle time must be given to that direction.
   $G_{TN1} \leftarrow VNR_{GN1} * $ Full_Cycle_Time (120 Seconds)
   $G_{TN2} \leftarrow VNR_{GN2} * $ Full_Cycle_Time (120 Seconds)

7. **DETERMINE** the next phase time.
   Next_Phase_Time $\leftarrow$ Average ($G_{TN1}, G_{TN2}$)

Fig.12. DT3P Next Phase Time Decision Making Algorithm

## V. Simulation Results

The experiments have been designed to show how efficient the DT3P is in terms of balancing the queue lengths on an intersection's legs, compares to the four other methods; two bench mark methods (Fixed and Actuated methods) and two latest methods ([6]'s and [1]'s methods). All the experiments would be started at the moment follows solving an emergency case, as illustrated before.

In Fig.13, The response of the five methods to the initial queues is shown as the experiments were run for four cycles time. As can be seen, both of the Fixed and the Actuated controllers have no response to the queue changes. While all the other three methods have responded and have changed the amount of the green time given for each direction as required but with different ratios.

After running the same setup for one hour, it has been obvious how efficient the DT3P's decisions compares to all the other methods. In Fig.14, DT3P tops the rest of the methods in terms utilizing the given green time. In other words, DT3P has made more accurate decisions that helped to decrease the waste of time. In addition to that, by looking to Fig.15, DT3P decisions have helped to decrease the average queue lengths at the intersection. The DT3P curve shows that it is the nearest to a straight line which means the intersection's situation have become more stable compares to the rest of the methods.

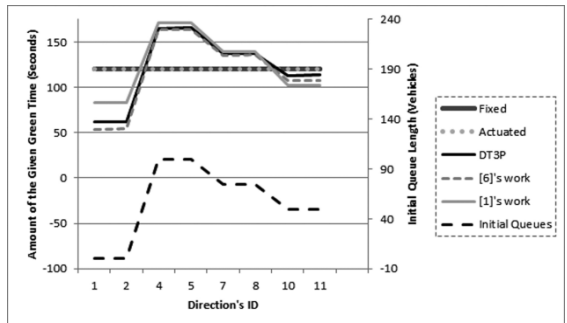

Fig.13: Short term System behavior; The Amount of the Given Green Time (Primary Axis) for each Direction by each of the five Methods

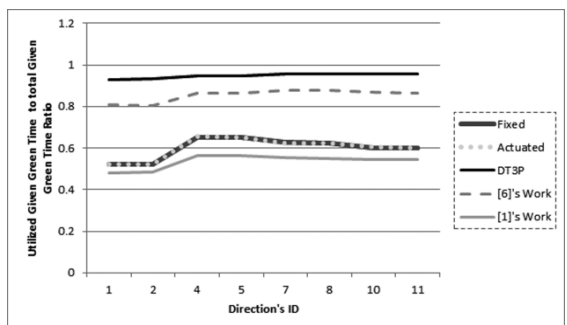

Fig.14: Long Term experimental result for the given green time utilization ratio

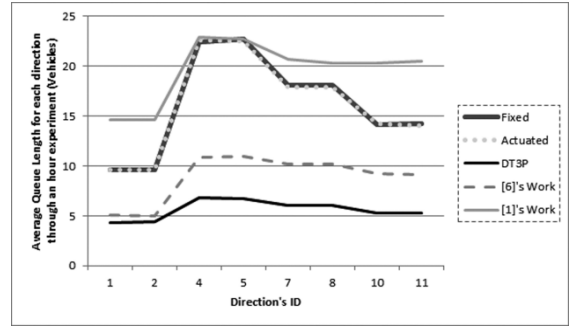

Fig.15. Average Queue Lengths for all the intersection's eight directions.

## VI. Conclusion

Many researchers have suggested and patented emergency vehicles detection systems however, according to our knowledge, none of them considered solving the effect of giving extra green time to a road while the queues are being built on others. This paper considers the problem of finding a better traffic light phase plan to stabilize/recover the situation at an effected intersection after solving an emergency vehicle existence. Logically, the higher the green time utilization is, the less average queue length would be. DT3P considered giving more accurate phase plan which led to have much better green time utilization which finally led to achieve less average queue length. The only limitation for this approach is its cost which high as it is needed to setup extra devices (RSEs and Vehicles detection belts). But an important point to be highlighted is that the suggested hardware setup can be used for other VANET applications in addition to the DT3P, which makes it cost-worthy.

## Acknowledgements

This work was supported by Universiti Teknologi Petronas under the Graduate Assistantship scheme.

M. K. Abbas, M. N. Karsiti, M. Napiah, S. Brahim

# Authors' information

[1] Electrical and Electronics Engineering Department, Universiti Teknologi Petronas.

[2] Electrical and Electronics Engineering Department, Universiti Teknologi Petronas.

[3] Civil Engineering Department, Universiti Teknologi Petronas,

[4] Mathematics and computer science department, College of science, Al-Faisal University

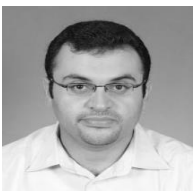

**Maythem Kamal Abbas** biographies was born in Baghdad, Iraq, in 1984. He got his BSc in Control and Computer Systems Engineering from University of Technology in Baghdad, Iraq in 2005. Subsequently he completed his Higher Diploma in IT in Bickenhall College of Computing (Middlesex University partner) in London, United Kingdom in 2008. Consequently, he completed his Master's degree in IT from Universiti Teknologi Petronas, Malaysia in 2009. Since then, he is doing his PhD in Electrical and Electronic Engineering in Universiti Teknologi Petronas.

Since 2008, he is a researcher with Information Technology Department and Electrical and Electronics Engineering department in Universitit teknologi Petronas. He is the author of 10 conference publications and 3 Islamic-teachings books (In Arabic Language). His research interests include vehicular ad hoc networks, formal specification language, decision making algorithms, protocol design, intelligent systems and robotics. He served as a reviewer and as a technical programme committee in many international conferences.

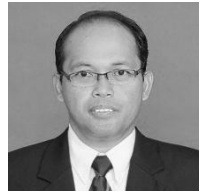

**Mohd Noh Karsiti** completed his training and obtained the degrees of Bachelor of Science in Electrical Engineering and Masters of Science in Electrical Engineering from California State University, Long Beach, USA, in 1985 and 1987 respectively. Subsequently, he completed his doctoral program in University of California, Irvine and was awarded PhD in Electrical and Computer Engineering in 1991.

From 1989 till 1999 he was a lecturer and Program Chair for Control Systems at the School of Electrical and Electronic Engineering, Universiti Sains Malaysia. He served USM for almost 10 years. He joined UTP in 1999 as a lecturer and later promoted to the current position of Associate Professor in 2005. He has been responsible for the development of the EEE program including curriculum design, facility developent, quality assurance, student advisory, staff training & development and R&D. From 2001 to 2006 he served as the Head of Department for Electrical and Electronic Engineering and currently he is an Associate Professor at the Department of Electrical and Electronic Engineering, Universiti Teknologi PETRONAS, Malaysia. He had served at various capacities including the Dean of Centre for Graduate Studies from 2007 to 2012.

He has taught many courses ranging from Control Systems, Data Communication Networks, Circuit Theory and even Computer programming. He is also actively involved in research activities. He has presented and published numerous technical publications in conferences and refereed journals. His current research interest spans from Intelligent System, Control, Robotics, Hybrid Vehicle, Process Modelling, and Tomography.

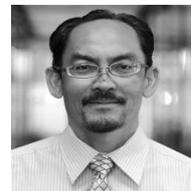

Assoc.Prof.Dr. Madzlan Napiah is a lecturer in the Civil Engineering Department Universiti Teknologi PETRONAS. He holds a PhD in Civil Engineering and MSc in Transportation Planning & Engineering, both from the University of Leeds UK. He also holds a BSc in Civil Engineering from Michigan State University USA. His research interests are in modelling and optimisation of public transportation, traffic flow, and also in highway materials and project planning and management.

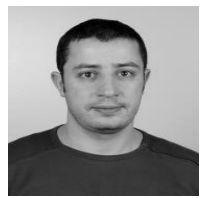

Samir Brahim-Belhaouari received the Master's degree in Telecommunications in 2000 from Institut Nationale Polytechnique (INP) of Toulouse, France and the PhD in 2006 from the Federal Polytechnic School of Lausanne (EPFL). He has been Senior Lecturer at Universiti Teknologi PETRONAS since 2007. His research interests lie in the areas of image and speech processing and stochastic processes.